\documentclass[preprint]{aastex}
\usepackage{amsmath}

\newcommand{\el}{{\rm e}}
\newcommand{\xel}{x_{\rm e}}

\newcommand{\h}{{\rm H}}
\newcommand{\mh}{m_{\rm H}}
\newcommand{\xh}{x({\rm H})}
\newcommand{\nh}{n_{\rm H}}

\newcommand{\hp}{{\rm H}^+}

\newcommand{\xhe}{x({\rm He})}

\newcommand{\hm}{{\rm H}_2}
\newcommand{\xhm}{x({\rm H}_2)}

\newcommand{\hmp}{{\rm H}_{2}^{+}}

\newcommand{\hthreep}{{\rm H}_{3}^{+}}

\newcommand{\atomicc}{{\rm C}}

\newcommand{\cp}{{\rm C}^+}

\newcommand{\ntwo}{{\rm N}_{2}}

\newcommand{\oh}{{\rm OH}}

\newcommand{\htwoo}{{\rm H}_{2}{\rm O}}
\newcommand{\xhtwoo}{x({\rm H}_{2}{\rm O})}

\newcommand{\co}{{\rm CO}}

\newcommand{\hcop}{{\rm HCO}^+}

\newcommand{\ch}{{\rm CH}}
\newcommand{\chp}{{\rm CH}^+}
\newcommand{\chtwo}{{\rm CH}_2}
\newcommand{\chtwop}{{\rm CH}_2^+}
\newcommand{\chthree}{{\rm CH}_3}
\newcommand{\chthreep}{{\rm CH}_3^+}
\newcommand{\chfour}{{\rm CH}_4}

\newcommand{\Td}{T_{\rm d}}
\newcommand{\Tg}{T_{\rm g}}
\newcommand{\nd}{n_{\rm d}}

\newcommand{\phchemheat}{{\rm photochemical \; heating}}
\newcommand{\Msun}{M_{\odot}}

\newcommand{\pcc}{{\rm cm}^{-3}}

\newcommand{\sqcm}{{\rm cm}^{2}}

\newcommand{\ps}{{\rm s}^{-1}}
\newcommand{\ccps}{{\rm cm}^{3} {\rm s}^{-1}}

\newcommand{\ev}{{\rm eV}}
\newcommand{\ergps}{{\rm erg}\,{\rm s}^{-1}}

\newcommand{\bcen}{\begin{center}}
\newcommand{\ecen}{\end{center}}
\newcommand{\be}{\begin{equation}}
\newcommand{\ee}{\end{equation}}
\newcommand{\bdis}{\begin{displaymath}}
\newcommand{\edis}{\end{displaymath}}

\def\noi{{\noindent}}
\def\ra{{\; \rightarrow}\;}

\slugcomment{July 9, 2015}

\begin{document}

%\title{Draft for Photochemical Heating}
\title{Photochemical Heating of Dense Molecular Gas}

\author{A. E. Glassgold}
\affil{Astronomy Department, University of California, Berkeley, CA 94720} 

\and 

\author{J. R. Najita}
\affil{National Optical Astronomy Observatory, 950 North Cherry Avenue, Tucson, AZ 85719}

%# LaTeX problem with next command
%\begin{abstract}

{\bf Abstract} - 
Photochemical heating is analyzed with emphasis on the heating generated by chemical reactions initiated by the products of photodissociation and photoionization. The immediate products are slowed down by collisions with the ambient gas and heat the gas. In addition to this direct process, heating is also produced 
by the subsequent chemical reactions initiated by these products. Some of this {\it chemical} heating comes from the kinetic energy of the reaction products and the rest from collisional de-excitation of the product atoms and molecules. In considering dense gas dominated by molecular hydrogen, we find that the chemical heating is sometimes as large if not much larger than the direct heating. In very dense gas the total photochemical heating approaches 10\,eV per photodissociation (or photoionization), competitive with other ways of heating molecular gas.    

\noi {\it Keywords} - protoplanetary disks, T Tauri stars

\section{Introduction}

Photochemical heating is important in dense photon transition regions (PDRs), especially where the gas changes from atomic to molecular. There are two main types of photochemical heating: "direct", coming from the kinetic energy of the immediate products of photoionization or photodissociation, and "chemical", arising from exothermic reactions of the products with abundant neutral species. The latter has received much less attention than the former, partly because it is sensitive to the detailed physical conditions of the gas, i.e. to density, temperature and chemistry. In this paper we develop photochemical heating for the dense and warm inner regions of protoplanetary disks and treat direct and chemical heating on an equal footing. We also discuss how the results may apply generally to diffuse and dense interstellar gas. 

Early consideration of $\phchemheat$ arose in the photoionization of cool,  diffuse interstellar gas (Spitzer 1978). Thus photoionization of the carbon atom imparts a small amount of kinetic energy to the products, a $\cp$ ion and an  electron.  Recombination then leads to escaping radiation and the net heating per ionization is small, $\simeq 1$\,eV per photoionization. Spitzer mentioned the possibility of chemical heating from the reaction of $\cp$ with molecular hydrogen (Dalgarno \& Oppenheimer 1974). This is not important for cool, diffuse gas because $\cp$ interacts weakly with $\hm$ by radiative association, and also because it depends on the abundance of $\hm$ which may be small in this case.

Henry \& McElroy (1968) treated chemical heating in connection with the photoionization of $\hm$ by photons with $h\nu > 15.44$\,eV from solar EUV irradiation of Jupiter's upper atmosphere. They examined the thermal consequences of the reactions of $\hmp$, e.g., 
\be
\label{h2+}
\hmp + \hm \rightarrow \hthreep + \el, \hspace{0.5in}
\hthreep + \el \rightarrow \hm + \h, 
\ee
and obtained chemical heating of order 10\,eV per ionization of $\hm$. These same processes were used by Glassgold \& Langer (1973) and Cravens \& Dalgarno (1978) for cosmic ray heating of interstellar molecular gas. A comprehensive treatment of heating by cosmic rays and X-rays, based on the analysis of electron interactions in an H, $\hm$ and He gas mixture by Dalgarno, Yan \& Liu (1999), has been given by Glassgold, Galli \& Padovani (2012; henceforth GGP12).

Chemical heating was actively developed for heating inner cometary comae from the photodissociation of $\htwoo$ by solar FUV and EUV radiation\footnote{In this paper we use the Lyman limit $\lambda = 911.7$\,\AA\ to distinguish EUV from FUV radiation.} (Marconi \& Mendis 1982; Ip 1983; Crovisier 1984, 1989; Rodgers \& Charnley 2002). Some of these papers focused on direct heating, but Marconi \& Mendis also included chemical heating. Some level of $\phchemheat$ has been included by Gorti \& Hollenbach in modeling protoplanetary disks (Gorti \& Hollenbach 2004, 2009, 2011) but with few details. Woitke et al.~(2009) and Woods \& Willacy (2009) included $\cp$ and $\hm$ photochemical heating in related studies. 

In this paper we treat both direct and chemical heating of dense molecular gas dominated by molecular hydrogen and exposed to FUV radiation. Although the inner regions of protoplanetary disk atmospheres provide the prime motivation, we also discuss molecular  clouds and PDRs. We will find that chemical heating can be at least the same order as the direct heating, and in some cases much larger. The latter situation tends to occur for targets like $\hm$ and CO, whose dissociation requires FUV photons with energies greater than 11\,eV. Some of the chemical heating is extracted from the chemical energy of the neutrals with which the dissociation products interact, or which they generate. 

When photodissociation or photoionization occurs, the incident photon energy is converted 
into kinetic energy of the products, and that leads to direct heating. The remainder goes into a 
product radical or ion with greater chemical energy that can engage in reactions that generate more 
heating. The photoionization of carbon provides a simple example, $h\nu + {\rm C} \, \ra \, \cp + \el$. 
Most of the kinetic energy goes to the electron, and the other product is the reactive $\cp$ ion. In 
the dense molecular regions of interest here, it will react with $\hm$ or other species rather than 
recombine radiatively, and these reactions can lead to further heating. Some of the species in the 
initial photo process or in subsequent reactions may be produced in excited levels, and some of this 
excitation may also be extracted as heat by collisional de-excitation of abundant species like H, 
$\hm$ and He if the density is high enough. In some cases, more energy can be obtained as heat than 
possessed by the initiating photon. This somewhat surprising outcome is a consequence of the fact that 
photodissociation regions are not in complete thermodynamic chemical equilibrium. They contain 
atoms and radicals with significant chemical energies that can be tapped by exothermic chemical 
reactions initiated by either the primary or the secondary products of the initial photodissociation 
or photoionization. This process obeys the  laws of total energy and mass conservation, which are 
automatically obeyed in our treatment of chemical reactions and their energetics. The chemical energy 
of the reactants is potential energy in the usual sense, i.e., electrostatic energy of ions and electrons 
in the atoms and molecules of the gas. Additional heating results when some of the chemical energy 
gets converted into kinetic energy while conserving the total energy. A particularly interesting 
example is the heating that results when the formation on dust grains leads to excited $\hm$ molecules, 
whose excitation can be converted to heat by collisional de-excitation at high densities. This possibility 
arises because molecular dissociation usually leads either directly or indirectly to H atoms, as will 
be discussed at length in Section 3.

Here our main goal is to call attention to the potentially significant role of photochemical heating in photodissociation regions. To illustrate this process, we discuss a number of simplified examples and give qualitative numerical estimates which would need to be refined for  specific applications. In the next section we give a short general development and then take up specific cases in Section 3. The significance of the results are discussed in Section 4 and in the brief conclusion in Section 5.

\section{Method}

We consider the $\phchemheat$ of slightly ionized molecular gas by FUV radiation with wavelengths 
$\lambda > 911.7$\,\AA. The heating rate per unit volume is expressed as,  
\be
\label{heating_general}
\Gamma_{\rm phchem} = \Gamma_{\rm dir} + \Gamma_{\rm chem}, 
\ee
where the ``direct'' heating  $\Gamma_{\rm dir}$ comes from the kinetic energy of the dissociation (or ionization) products and the chemical heating 
$\Gamma_{\rm chem}$ comes from their chemical reactions. Each of the terms in 
Eq.~\ref{heating_general} is a sum over neutral species X of the form,
\be
\label{heating_specific}
\Gamma({\rm X}) = G({\rm X})\,  n({\rm X})\, Q({\rm X}), 
\ee
where $G({\rm X})$ is the photodissociation or photoionization rate of X with volumetric density $n({\rm X})$ and 
$Q({\rm X})$ is the heating energy, either direct or chemical, per photodissociation 
(or photoionization)\footnote{In principle, Eq.~\ref{heating_specific} applies to molecular ions but they are usually 
less important for photochemical heating than the neutrals.}. Atomic species such as neutral carbon can contribute as 
well as molecules, and then $G({\rm X})$ is the photoionization rate. Equation~\ref{heating_general} would appear as 
one term in the heat equation of a full thermal-chemical model that would also include a system of chemical equations 
as well as equations for the dynamics and radiative transfer. We focus here on the microscopic physics 
and chemistry of the photochemical heating of selected molecules and make qualitative estimates for these species. 
These results potentially apply to both time-varying and steady astrophysical situations.  We do not incorporate 
the findings in a general thermal-chemical model in this paper, but provide several example applications to protoplanetary disks. 

The absorption of a FUV photon may result in fluorescence rather than dissociation. This can lead to heating when the final molecule's low-lying ro-vibrational levels are collisionally de-excited. The primary example is molecular hydrogen 
(e.g., Tielens 2005, Sec.~3.4; Woitke et al.~2009). In this case, $\Gamma_{\rm dir} = 0$, $G({\rm X})$ is the photo-excitation rate and $Q({\rm X})$ is the photo-excitation heating per excitation derived from collisional de-excitation. The most important case of $\hm$ is discussed in Sec. 3.2. 

The chemical reactions initiated by photodissociation also produce species in excited states that can lead to 
heating following collisional dissociation. This excitation energy is not always negligible, and it introduces 
a significant dependence on density due to variations in the critical density from transition to transition. 
The critical density for a transition $u \ra l$ is $n^{\rm cr}_{u,l} = A_{u,l} /  k_{u,l}$, where $A_{u,l}$ is
the effective $A$-value of the transition and $k_{u,l}$ is the rate coefficient for collisional de-excitation. 
Depending on the nature of the transition, both $A_{u,l}$ and  $k_{u,l}$ can vary widely. Again $\hm$ provides 
an important example, where the critical densities of the pure rotational transitions are  $\sim 10^2 - 10^3 \pcc$ 
for modest excitation temperatures whereas those for ro-vibrational transitions are many orders of magnitude larger, 
$\sim 10^{10} \pcc$ (Le Bourlot, Pineau des For\^ets \& Flower 1999). Taking into account the possible large 
differences in excitation energy, the resulting chemical heating is very sensitive to density. Indeed, the small scale of the pure rotational energy of $\hm$ compared to the level of chemical heating (tenths vs.~several eV) 
means that the former can usually be ignored.  

In the next section we estimate the heating energy $Q({\rm X})$ for some abundant neutrals X in interstellar 
and circumstellar matter. We assume that the FUV photo-rate $G({\rm X})$ and the run of density $n({\rm X})$ 
are known for each application. Not only is the photo-rate sensitive to the details of the radiation source, 
it is altered by attenuation and scattering, which are sensitive to composition. Beyond the first two factors 
in Eq.~\ref{heating_specific}, the heating energy also depends on the temperature and chemical composition 
of the gas. In a self-consistent model of the gas, the photochemical heating rates in 
Eq.~\ref{heating_specific} would be included in the heat equation and thus affect the density and temperature.
We do not attempt to make such complete models here but only estimate the rough magnitudes of the heating rates for two extreme density regimes.  We discuss but do not treat in detail gas 
heating due to the photo-electric effect on dust particles (Tielens 2005, Sec.~3.3) or the heating by cosmic ray 
or X-ray generated electrons (GGP12), which are both treated in the cited references. 
In general, dense gas is dissociated and ionized by FUV, X-rays and cosmic rays. Model calculations of gas thermal 
properties should include the heating from all of these sources, as developed here and in GGP12. Other than the 
formation of molecular hydrogen, the entire analysis in this paper is based on gas phase chemistry.

The approach followed in this paper is based on previous experience in modeling PDRs and protoplanetary 
disks which helps identify the most abundant species in the molecular regions of interest. The detailed 
estimates for specific cases that follow should serve as paradigms for the calculation of photochemical heating for situations not treated here. An alternate approach was introduced by Clavel, Viala \& Bel (1978) and recently used by Du \& Bergin (2014) for protoplanetary disks. A similar methodology has been employed in PDR codes (Le Petit et al.~2006; Rollig et al.~2007).  Clavel et al. simply added up all of the energy yields for the exothermic reactions in their thermal-chemical model of dark clouds. They pointed out that this method is prone to over-estimating the chemical energy because it assumes that all of the energy yield goes into the kinetic energy of the products of chemical reactions and thus into heating. Indeed much of the following discussion deals with the extent to which the available energy goes into heating, as opposed to excitation and fluorescence. The special focus of the present paper is on the heating associated with the absorption of external FUV radiation.

\section{Photochemical Heating of Abundant Species}

Photodissociation and chemical heating depend on the electronic structure of the individual species. In this section 
we show how the photochemical heating is obtained for some of the more important species that arise in applications 
to interstellar and circumstellar matter. In accord with Eq.~\ref{heating_specific}, photochemical heating is important 
only where the the relevant FUV flux and the density of the absorber are significant. The examples discussed here were 
chosen because they are operative in  the inner regions of protoplanetary disk atmospheres. We start with atomic carbon, 
which at first sight might seem to be the simplest case. However, unlike the other molecular species treated in this 
section, it is more complicated because of the enhanced reactivity of $\cp$ as compared to the neutral fragments mainly 
generated by molecular photodissociation. The case of atomic carbon illustrates another issue, which is that it may not 
be very abundant where the reactants exist that can lead to photochemical heating. In photodissociation regions, C is 
transformed into CO, often passing first through $\cp$. The formation of CO depends on the formation of $\hm$, and $\hm$, 
CO and C are all destroyed by photons in the same FUV band, with $\hm$ eventually generating most of the attenuation. 
In transition regions from atomic to molecular, photoionization of C may occur in the presence of $\hm$, and this indeed 
occurs in the inner atmospheres of protoplanetary disks. But even this case is complicated because heavier molecules may 
also form if the disk is sufficiently warm, and then $\cp$ is destroyed by other molecules such as $\htwoo$ as well as $\hm$. 

\subsection{Atomic Carbon}

With an ionization potential $IP = 11.26$\,eV, the ionization threshold of atomic carbon is 1101\,\AA. The photoionization ionization cross section varies by no more 10\% in the interval between 911.7 and 1101\,\AA\ (Canto et al.~1981) with an average value of $1.58 \times 10^{-17}\, \sqcm$. The energy generated by the reaction,
\be 
\label{Cphion}
h\nu + {\rm C} \, \ra \, \cp + \el
\ee 
is
\be
\Delta E = h\nu + E_{\rm excn}({\rm C}) - E_{\rm excn}(\cp) - IP({\rm C}),
\ee
where $E_{\rm excn}({\rm C})$ and $E_{\rm excn}(\cp)$ are the excitation energies of the initial atom and the final ion. For the moderate temperature molecular environments under consideration, they are just the excitations of the ground fine-structure levels of C and $\cp$ and no more than 0.00743 and 0.0165\,eV, respectively. Such small quantities can be ignored in considering heating energies of order 1\,eV or more. Thus a good estimate of the direct heating energy is,
\be
Q_{\rm dir}({\rm  C}) =  h\bar{\nu} -IP({\rm C}),
\ee
where $\bar{\nu}$ is the spectrum- and cross section-weighted average frequency. Assuming for purposes of estimation that the incident FUV radiation is roughly constant in the 911.7 - 1101\,\AA\, band, and recalling that the cross section is almost constant, the mean photon energy is 12.4\,eV and the direct photochemical heating for atomic C is, 
\be
\label{Cdirectheating}
Q_{\rm dir}({\rm  C}) = 1.14\, {\rm eV}.
\ee
This estimate applies to both diffuse and dense gas. It is often cited as the heating from the photoionization of C in neutral gas (e.g., Tielens, Sec.~3,2), but it is only a small part of the story, as shown below.

The chemical heating of $\cp$ depends on the abundance of the species with which it interacts. In molecular regions, $\hm$ is the obvious candidate, but the relevant rate coefficients are small at low temperatures\footnote{A list of two-body reactions is given in Table A1 at the end of the paper along with their maximum energy yields obtained from Table A2, which gives the chemical energies of the relevant species.}, i.e.,
\be
\label{R:cpreac1}
k(\cp + \hm \ra \chp + \h ) = 
7.5 \times 10^{-10}\,e^{-4620/T}\, \ccps,		\tag{R1}
\ee
and
\be
\label{R:cpreac2}
k(\cp + \hm \ra \chtwop + h\nu) \sim 
4.0 \times 10^{-16}\; \ccps,				\tag{R2}
\ee
where $k$ stands for rate coefficient (units $\ccps$). Reaction~\ref{R:cpreac1} is slightly endothermic, whereas Reaction~\ref{R:cpreac2} is exothermic, but its rate coefficient is small because  it is a radiative process. The value for \ref{R:cpreac1} is an approximate fit to the experimental values determined by Herr\'aez-Aguilar et al.~(2014). Oxygen molecules such as water are readily formed by radical reactions in warm regions with $T > 300$\,K, and we will also estimate the chemical heating from the reaction,
\be
\label{R:cpreac3}
k(\cp + \htwoo \ra \hcop + \h) =
2.1 \times 10^{-9}\; \ccps. 				\tag{R3}
\ee
At temperatures near 1000\,K, characteristic of the molecular transition region in protoplanetary disks at small radii, Reactions~\ref{R:cpreac1} and \ref{R:cpreac3} are both important, and we calculate their chemical heating for warm ($> 300$\,K) and strongly molecular regions ($\xhm \gg \xh$). If $\htwoo$ is not present or has a low 
abundance, as in cool for interstellar clouds, then the reaction of $\cp$ with $\htwoo$ can be ignored. 

\subsubsection{Chemical Heating of $\cp$ Reacting with $\hm$}

Following the initiating reaction~\ref{R:cpreac1}, successive fast exothermic reactions with $\hm$ lead to 
$\chthreep$. The next $\hm$ hydrogenation reaction is endothermic by $\sim 2.6$\,eV, and $\chthreep$ is destroyed 
by fast dissociative recombination with electrons,
\be
\label{R:cpreac4}
k(\el + \chthreep \ra {\rm products}) = 2.0 \times 10^{-5} \, T^{-0.61} \; \ccps. 	\tag{R4}
\ee
The products and branching ratios (Thomas et al.~2012) are given in Table 1.
\begin{center}
Table 1. Dissociative Recombination of $\chthreep$
\begin{tabular}{clc}
\hline
Branch & Products					& Branching Ratio\,(\%)				\\
\hline
1 & $\chthreep + \el \ra \chtwo + \h 		$	&	35				\\
2 & $\chthreep + \el \ra \ch + \hm 		$	&	10				\\
3 & $\chthreep + \el \ra \ch + 2\h 		$	&	20				\\
4 & $\chthreep + \el \ra \atomicc + \hm + \h  	$	&	35				\\
\hline
\hline
\end{tabular}
\end{center}
The $\ch$ radicals produced in branches 2 and 3 are converted into $\chtwo$ by the neutral reaction 
with $\hm$ with rate coefficient,
\be
\label{R:ch+h2}
k(\ch + \hm \ra  \chtwo +\h) =
2.9 \times 10^{-10 } \, e^{-1670/T}\, \ccps.	\tag{R5}
\ee
The rate coefficient for the reaction of $\chtwo$ with $\hm$ is very small, and instead 
$\chtwo$ interacts with atomic oxygen to produce CO,
\be
\label{R:O+CH2-CO+2H}
k({\rm O} + \chtwo \ra \co + 2\h) = 2.0 \times 10^{-10}\, e^{-270/T} \ccps,	\tag{R6a}
\ee
\be
\label{R:O+CH2-CO+H2}
k({\rm O} + \chtwo \ra \co + \hm) = 1.4 \times 10^{-10}\, e^{-270/T} \ccps.	\tag{R6b}
\ee

The overall effect of all of these reactions that follow from the photoionization of atomic C can be 
assessed by adding all of the contributing reactions, i.e., Eq.~\ref{Cphion}, Reaction~\ref{R:cpreac1}, the 
three fast ion-hydrogenation reactions ($\ch_n^+ + \hm \ra  \ch_{n+1}^+ +\h, \; n=0-2$), the $\chthreep$ 
dissociative recombination Reaction~\ref{R:cpreac4} with branching ratios in Table~1, and 
Reactions~\ref{R:O+CH2-CO+2H} and \ref{R:O+CH2-CO+H2} for the production of CO. Table~2 lists the net energy production from all of these reactions for each dissociative recombination branch. Each entry conserves 
energy and species number. There are now seven rows 
instead of four because the formation of CO (reactions~\ref{R:O+CH2-CO+2H} and \ref{R:O+CH2-CO+2H}) 
has two outcomes with weights of 60\% and 40\%. The fourth column gives the total available energy, 
$\Delta E$, including the direct heating of 1.14\,eV. The remainder goes into heating and excitation.  Some of the branch energies exceed the mean incident photon energy of 12.4\,eV, as expected on the basis of the discussion in Sec.~1. 
\begin{center}
Table 2. Energy Yields From the Photoionization of Atomic Carbon					\\
\begin{tabular}{clccc}    
\hline
Branch & Consolidated Reactions	& Ratio (\%)    & $ \Delta E$(eV) & 	$N(\h)^{1}$ \\
\hline
1a & $h\nu + {\rm C} + 3\hm + {\rm O} \ra \co + 6\h$ 	& 21	&  10.0	&	6	\\
1b & $h\nu + {\rm C} + 2\hm + {\rm O} \ra \co + 4\h$		& 14	&  14.5	&	4	\\
2a & $h\nu + {\rm C} + 3\hm + {\rm O} \ra \co + 6\h$ 	&  6	&  10.0	&	6	\\
2b & $h\nu + {\rm C} + 2\hm + {\rm O} \ra \co + 4\h$		&  4	&  14.5	&	4	\\
3a & $h\nu + {\rm C} + 4\hm + {\rm O} \ra \co + 8\h$ 	& 12	&   5.5	&	8	\\
3b & $h\nu + {\rm C} + 3\hm + {\rm O} \ra \co + 6\h$		&  8	&  10.0	&	6	\\
4  & $h\nu + {\rm C} + 2\hm      \ra \atomicc + 4\h$	 	& 35	&   3.4	& 	4	\\
\hline
   & Averages 						&    	&    8.0 &      5	 \\
\hline
\multicolumn{5}{l}{1. $N(\h)$ is the number of H atoms generated in each branch.} 
\end{tabular}
\end{center}

The last column in Table 2 is the number of H atoms produced in each branch from the chemical 
reactions generated by the $\cp$ ion in the photoionization of atomic C. There are implications 
for gas heating from the energy generated in the formation of $\hm$ on grains,
\be
\label{R:H2formation}
\h + \h + {\rm Gr} \ra \hm + {\rm Gr}^{*}.			\tag{R7}
\ee
The symbol ${\rm Gr}^{*}$ indicates that the grain is left excited in the $\hm$ formation process. The 
rest of the 4.5\,eV binding energy goes into the kinetic and internal excitation energy of the newly 
formed molecule. Some of this energy can heat the gas and is often referred to as $\hm$ formation heating, 
but the partitioning into $\hm$ kinetic and excitation energy and grain excitation is poorly understood. 
A frequent assumption is that there is equipartition among the three possibilities. However, on
the basis of laboratory experiments by Lemaire et al.~(2010), we have modeled protoplanetary disks 
assuming $\sim 30\%$ of the binding energy ($\sim 1.3$\,eV) goes into the rovibrational excitation and very 
little into kinetic energy (Adamkovics, Glassgold  \& Najita, 2014; henceforth AGN14). Recovery 
of the excitation energy as heat requires high densities to collisionally de-excite the molecules, 
$ \leq 10^6 \pcc$ for pure rotational transitions and much higher densities for vibrational excitation,  
$10^8 - 10^{11} \pcc$ (Le Bourlot et al.~1999). By way of comparison, the Meudon PDR code 
uses twice this $\hm$ excitation energy and a small amount of kinetic energy $\sim 0.6$\,eV (Le Bourlot et al.~2012).

If $ \Delta E(\hm)_{\rm form}$ is the heating from the formation of an $\hm$ molecule, then each dissociation 
channel in Table~2 leads to an additional heating of $ N(\h) \Delta E(\hm)_{\rm form}$, where $ N(\h)$ is the 
number of H atoms generated given in the last column of Table 1.  Each photochemically produced H atom 
combines with the required second atom on a grain surface. With an average value of $ N(\h) = 5$, $\hm$ 
formation heating is comparable to the chemical energy in the fourth column of Table~2, which does not include 
$\hm$ formation heating. $\hm$ formation heating is usually included in thermal chemical calculations of 
interstellar and circumstellar matter as a general heating process; it is not explicitly tabulated here 
to avoid duplication with the usual treatment. It is of course important to remember that 
$\Delta E(\hm)_{\rm form}$ depends on the density of the gas and on the properties of the grain surface, 
particularly on the assumption that the grains have atomic hydrogen accessible for molecule formation. 
The kinetic part is generally available in any $\hm$ region, whereas the $\hm$ excitation energy needs 
high densities to be converted into heating. One case where this distinction vanishes is the very high 
density regions of protoplanetary disks. However the exact value of $\hm$ formation heating remains  
uncertain, as discussed for example by Bechellerie et al.~(2008) and Le Bourlot et al.~(2012). 

Table 2 lists 8.0\,eV as the mean energy available following the photoionization of a C atom and leading mainly 
to the formation of CO. Not all of this is available for heating, however, except possibly at very high densities. Because reaction~\ref{R:cpreac1} is slightly endothermic and the subsequent hydrogenation reactions slightly exothermic, the energy produced just to reach $\chthreep$ via $h\nu + {\rm C} + 3\hm \ra \chthreep + 3\h $
is only 1.7\,eV. Of this, 1.1\,eV goes into the kinetic energy of the photoelectron, and 0.6\,eV is shared between the kinetic energy of the H atoms from the hydrogenation and excitation of the hydrocarbon ions CH$_n^{+}$ ($n=1-3$). 
Assuming equal sharing between kinetic and excitation energy, the heating generated by forming 
$\chthreep$ is $\simeq 1.4$\,eV. 

The dissociative recombination of the $\chthreep$ ion provides energetic H atoms and $\hm$ molecules 
that can heat the gas directly. 
Other than the branching ratios, however, there is very little experimental information on the 
energetics of the dissociative recombination of $\chthreep$, except for the fourth branch (with an 
available energy of 3.4\,eV), where half the time the C atom is produced in the $^1$D$_2$ level at 
1.26\,eV (Thomas et al.~(2012). For a handful of cases where storage ring experiments provide 
information on the energetics of the dissociation of heavy molecular ions, the results vary widely
(e.g., Thomas et al.~2005; R.~D.~Thomas, private communication 2015). 

We estimate that on average there is equal sharing between kinetic 
and excitation energy in both the production and recombination of $\chthreep$. Thus of the $\simeq 
1.7$\,eV available in the production of $\chthreep$ from $\cp$, direct heating accounts for 1.1\,eV 
and 0.6\,eV for heating and excitation, for an estimated heating of $\simeq 1.4$\,eV. Of the 6.3\,eV 
available from the dissociative recombination of $\chthreep$, $\simeq 3.1$\,eV might be available for 
heating. Adding in the 1.14\,eV in direct heating, there is an estimated 4.6\,eV in heating and an 
additional 3.4\,eV in excitation. 

To summarize, the heating made available from the photoionization of atomic C and the subsequent 
interaction of the $\cp$ ion with $\hm$ is in two limiting cases: (i) $\simeq 4.6$\,eV for 
{\it moderately dense $\hm$ regions},  i.e, those not dense enough for newly formed species to be 
collisionally de-excited; and (ii) $\simeq 8 $\,eV for {\it very high density regions}, where 
collisional de-excitation is effective. In both cases, the direct heating of 1.14\,eV has been included 
in these estimates.

\subsubsection{Heating of $\cp$ Reacting with $\htwoo$}

The primary reaction of $\cp$ with $\htwoo$ leads to $\hcop$ and ground state atomic H,
\be
\label{R:hco+prod}
k(\cp + \htwoo \ra \hcop + \h) = 2.1 \times 10^{-9} \, \ccps.		\tag{R3}
\ee 
This is followed by the dissociative recombination of $\hcop$, which 87\% of the time leads to
CO + H, and has the measured rate coefficient (Hamberg et al.~2014).
\be
\label{R:DRhcop+}
k(\el + \hcop \ra \co + \h) = 1.8\times 10^{-5}\, T^{-0.79}\, \ccps.	\tag{R8}.
\ee
The combined effect of Reactions~\ref{R:hco+prod} and \ref{R:DRhcop+} is,
\be
\label{hco+1}
\el + \cp + \htwoo \ra \co + 2\h,
\ee
which has a maximum energy yield of 12.7\,eV, 5.25\,eV from reaction~\ref{R:hco+prod} and 7.46\,eV from 
reaction~\ref{R:DRhcop+}. The heating from the formation of two $\hm$ molecules is not included 
here for the reason discussed in the previous subsection. Some fraction of the energy from Eq.~\ref{hco+1} 
may lead to excitation that can be recovered at very high densities. We estimate that 50\% of the 5.25\,eV 
available from reaction~\ref{R:hco+prod}, or 2.6\,eV, leads to heating at moderate densities with the
full amount becoming available at very high densities. 

The CO produced in the recombination of $\hcop$ may be in the ground level (X $^1\Sigma^+, \sim 50\%$) or 
the next two electronic levels (a $^3\Pi,\,6.0\ev; \sim 25\%$) and (a` $^3\Pi^+,\,6.3\ev; \sim 25\%$) according 
to Herman et al.(~2014). The excited levels are slow to decay to the ground level, e.g., $A(a \ra X) = 133\,\ps$, 
and collisional de-excitation requires $\hm$ densities greater than $10^{12} \, \pcc$. Thus only a small 
amount of energy is available for heating when the recombination of $\hcop$ is to excited electronic levels 
of CO, $\simeq 1.0$\,eV. On averaging over both outcomes of the dissociative recombination of $\hcop$, 
$\sim 4.2$\,eV is available for kinetic energy and excitation instead of 7.5\,eV. If equal amounts go into 
kinetic energy and excitation, then the recombination of $\hcop$ leads to heating of $\sim 2.1$\,eV at moderate densities and twice that at very high densities. 

To summarize, these estimates of $\cp$ reacting with $\htwoo$, lead to the following contributions to the
photochemical heating for (i) {\it moderately dense regions}: 1.1\,eV direct, 2.6\,eV from Eq.~\ref{R:hco+prod}, 
and  2.1\,eV from Eq.~\ref{R:DRhcop+} for a total of 5.8\,eV, and (ii) {\it very dense regions}: 1.1\,eV direct, 
5.2\,eV from Eq.~\ref{R:hco+prod}, and  4.2\,eV from Eq.~\ref{R:DRhcop+} for a total of 10.6\,eV.

\subsubsection{Summary of $\cp$ Chemical Heating}

The chemical heating per photoionization $Q_{\rm chem }({\rm C})$ depends on the probabilities and 
energy yields for $\cp$ reacting with $\hm$ or with $\htwoo$, i.e., 
$f(\cp + \hm)$ and  $f(\cp + \htwoo)$, and the corresponding energy yields, 
$\Delta E(\cp +\hm)$ and $\Delta E(\cp + \htwoo)$,  
\be
\label{cpnetheat}
Q_{\rm chem}(\cp) = 
f(\cp + \hm)\,\Delta E(\cp + \hm) + f(\cp + \htwoo)\,\Delta E(\cp + \htwoo).
\ee
The fractions depend depend on the corresponding rate coefficients and abundances,
\be
\label{cpfractions}
f(\cp + \hm) = \frac{k(R1) \, \xhm}
{k(R1) \, \xhm + k(R3) \, \xhtwoo},	
\hspace{0.10in}
f(\cp + \htwoo) = \frac{k(R3) \, \xhtwoo}
{k(R1) \, \xhm + k(R3) \, \xhtwoo}.
\ee
Because Reaction~\ref{R:cpreac1} has a substantial energy barrier, $f(\cp + \hm)$ is much less than $f(\cp + \htwoo)$ until temperatures as high as 600\,K are reached. Table 3 gives the estimates for $\Delta E(\cp + \hm)$ and $\Delta E(\cp + \htwoo)$ made in the preceding paragraphs in the two limits discussed there. Eq.~\ref{cpfractions} 
shows how the photochemical heating of atomic C via the reaction of $\cp$ and $\htwoo$ depends on the abundance of 
$\htwoo$ and the corresponding rate coefficient.  

\begin{center}
Table 3. Carbon Photochemical Heating in eV	\\
\begin{tabular}{ccc}    
\hline
Density 		& $\Delta E(\cp + \hm)$    & $\Delta E(\cp + \htwoo)$    \\
\hline
moderately dense	& 4.6			   & 5.8			\\
very dense		& 8.0			   & 10.6			\\
\hline	
%\multicolumn{3}{l}{}
\end{tabular}
\end{center} 
They  depend explicitly on the chemistry through the abundances of $\hm$ and $\htwoo$ and on the temperature 
through the rate coefficients. These numbers are significantly larger than the conventional direct heating in 
Eq.~\ref{Cdirectheating}, 1.14\,eV, which is included in all the entries in Table 3.  If reactions with other molecules 
are relevant, they would be treated in a similar manner.

\subsection{Molecular Hydrogen}

The photodissociation of $\hm$ follows the absorption of FUV photons into excited electronic levels that decay by fluorescence to the X$^1 \Sigma_u^+$ ground level 
($\sim 85$\,\%) and to the 2\,H($^2S_{1/2}$) continuum ($\sim 15$\,\%)). Most of the dissociation occurs from the B$^1 \Sigma_u^+$ level starting at 11.37\,eV above the ground X$^1 \Sigma_u^+$ level. Applying energy conservation to both the initial excitation reaction and the subsequent dissociation, we find that,
\be
\label{excndiss}
h\nu_{\rm in} + E_{\rm excn}({\rm X}) = 
h\nu_{\rm diss} + D(\hm) + \Delta E_{\rm kin},
\ee  
where $h\nu_{\rm in}$ is the energy of the exciting photon, $E_{\rm excn}({\rm X})$ is the 
initial (ro-vibrational) excitation of the target $\hm$ molecule, $h\nu_{\rm diss}$ is the 
energy of the photon emitted in the dissociation of $\hm$, $D(\hm) = 4.48$\,eV is the dissociation 
energy, and $\Delta E_{\rm kin}$ is the kinetic energy of the dissociated H atoms. 

Averaging over the incident photon energy distribution, we can use Eq.~\ref{excndiss} to  estimate ${\Delta E}_{\rm kin}$, the {\it mean} direct heating per photodissociation. Assuming that the incident FUV photon spectrum is approximately uniform over the 911.7\,-\,1100\,\AA\, band, the mean value of the dissociating photon energy is $h{\bar \nu}_{\rm in} = 12.5$\,eV, and the above equation becomes,
\be
\label{excndiss2}
h\nu_{\rm diss} + \Delta E_{\rm kin} - E_{\rm excn}({\rm X}) =
h{\bar \nu}_{\rm in} - D(\hm) = 8.0\,{\rm eV}. 
\ee
If the initial molecule is thermally excited with $T_{\rm excn} \sim 500-1000$\,K,  
$E_{\rm excn}({\rm X}) \sim 0.05-0.10$\,eV is a small and ignorable correction. The outgoing
photon energy $h\nu_{\rm diss}$ may be lower or higher than $E(B) - D(\hm) = 6.89$ \,eV, higher 
because the absorption may lead to an excited B state and lower because dissociation can lead 
to a range of continuum states above the dissociation limit. If we ignore these possibilities 
and approximate $h\nu_{\rm diss}$ by $E(B) - D(\hm) \simeq 6.9$\,eV, Eq.~\ref{excndiss2} yields 
$\Delta E_{\rm kin} \sim 1.2$\,eV, with considerable uncertainty, possibly of the same order 
of magnitude. This estimate of the direct heating on dissociation of the $\hm$ molecule 
assumes that the excitation energies before and after the photon is absorbed are negligible.

The heating associated with the photodissociation of $\hm$ is only a small 
part of the total heating because most of the FUV radiation absorbed by $\hm$ goes into fluorescent 
radiation from the B to the X ground level ($\sim$85\%), as opposed to dissociation ($\sim$15\%). 
Energy conservation applied to the dominant fluorescent mode reads,
\be
\label{excnfl}
h{\bar \nu}_{\rm in} + E_{\rm exin}({\rm X}) = 
h\nu_{\rm fl} + E_{\rm exfin}({\rm X}),
\ee 
where $E_{\rm exin}({\rm X})$ and $E_{\rm exfin}({\rm X})$ are, respectively, the 
initial and final average excitation energies and $h\nu_{\rm fl}$ is the average 
energy of the fluorescent photons from the B level, 
$h\nu_{\rm fl} \simeq E(B) - D(\hm)$ = 6.9\,eV. The heating is 
generated by collisional de-excitation of the ro-vibrational levels of the final 
X ground level, and is likely to be considerably smaller than 6.9\,eV.

Similar considerations of $\hm$ photo heating were {given in some detail in Sec~3.4 
of Tielens (2005) textbook with the same overall conclusion, i.e., heating from 
collisional de-excitation of ro-vibrational levels of the ground state associated with 
B to X fluorescent decay is the dominant mechanism.  We use his estimate for the 
excitation of the $\hm$ molecule following the decay of the transition from the B to 
the X level, $E_{\rm exfin}({\rm X}) = 2$\,eV, and assume that this can be recovered by 
collisional de-excitation at very high densities. Expressing this in terms of 
{\it photodissociation} rates by multiplying by 0.85/0.15, this becomes 11.5\,eV per 
photodissociation. Combining this estimate with the 1.2\,eV for direct photodissociation 
heating, the average total heating in dense regions is 12.5 per photodissociation.

\subsection{Carbon Monoxide}

Like $\hm$, the photodissociation of CO, 
\be 
\label{cophdiss}
h\nu + \co \ra {\rm C} + {\rm O}, 
\ee
proceeds by absorption of lines with wavelengths from 912 - 1076\, \AA \, or (11.5 - 13.6\,eV, van Dishoeck \& Black 1988; Visser et al. 2009). The high dissociation energy of CO (11.09\,eV) means that only modest amounts of the incident FUV energy are available for either direct heating or product excitation. The excited levels of CO are pre-dissociating and decay mainly into the ground levels of C and O. Those incident photons close to the Lyman limit could provide enough energy to excite the first forbidden levels of either C\,I or O\,I at 1.26 and 1.97\,eV, respectively, which would be collisionally de-excited at high density. However, we will ignore this possibility and assume that the small amount of energy available from Eq.~\ref{cophdiss} goes into direct heating.

The mean energy of the dissociating photons is 12.5\,eV so the energy yield 
of Eq.~\ref{cophdiss} is $\sim 1.4$\,eV, which we assume all goes into heating. This is a slight overestimate because we are ignoring the excitation of the C\,I and O\,I ground fine structure 
levels, whose upper levels are at 0.0053\,eV and 0.028\,eV respectively.

The dominant sources of the photochemical heating of CO are the chemical reactions initiated by the C and O dissociation products. For the O fragment, the dominant reactions in a dense, warm molecular regions are,
\be
\label{R:O+H2}
{\rm O} + \hm \ra \oh + \h,				\tag{R9}
\ee
and,
\be
\label{R:OH+H2}
\oh + \hm \ra \htwoo + \h.			 \tag{R10}
\ee
These two reactions are equivalent to,
\be
\label{Oheateq}
{\rm O} + 2\hm \ra \htwoo + 2\h, 
\ee 
with a net energy yield of 0.57\,eV. Some part of this may go into excitation of $\htwoo$ and OH and 
be converted into heat at high densities by collisional de-excitation. The maximum net chemical 
heating from the reactions of the O fragment is $\simeq 0.6$\,eV. The chemical heating from the C atom 
involves the reaction,
\be
\label{R:C+OH}
{\rm C} + \oh \ra \co + \h,			\tag{R11}
\ee
which has an energy release of 6.7 eV. 

To summarize, this analysis of the photochemical heating of CO gives a small amount of 
direct heating, $Q_{\rm dir}(\co) = 1.4$\,eV, and a large amount of chemical heating:  
$Q_{\rm chem }(\co) = 7.3$\,eV in very dense regions where collisional de-excitation 
of excited molecules is enable. In moderately dense regions, $Q_{\rm chem }(\co) \simeq 3.6$\,eV,
assuming that half of the available chemical energy is in kinetic energy. In addition, three H 
atoms are created and contribute to $\hm$ formation heating.

\subsection{Water}

The photodissociation cross section of water has been well measured (e.g., Lee \& Suto, 1986; Fillion et al.~2001; Parkinson \& Yoshino 2003; Mota et al.~2005); and summarized by Mota et al.~(2005) It is 
customary to divide the wavelength dependent cross section into two bands:
 
Band 1 from threshold (2475\,\AA) to $\simeq 1450$\,\AA \,

Band 2 from $\simeq 1450$\,\AA \,to the Lyman limit.

\noi The cross-section in Band 1 is basically smoothly varying, with  a broad maximum centered near 1650\, \AA . The measurements show that photodissociation in Band 1  yields directly the electronic ground states of OH and H,
\be
\label{h2o[hdiss1}
h\nu + \htwoo(\tilde{\rm X}) \ra \oh(\rm X) + \h(^2{\rm S}_{1/2})
\hspace{0.5in} ({\rm Band\; 1}). 		
\ee
According to the experiment of Andressan et al.~(1986), 88\% of the available energy goes into fragment kinetic energy, 10\% into vibrational excitation and 2\% into rotational energy of the ground level of OH. The last two could be available for a small amount of heating after collisional de-excitation.

Shortward of 1450\,\AA, the Band 2 cross section has small oscillations around a well defined continuum, 
although the oscillations became very large below 1250\,\AA. According to experiments by Harich et al.~(2000) with Ly$\alpha$ photons, there are three main channels for Band 2 photodissociation,
\be
\label{h2o[hdiss2a}
h\nu + \htwoo(\tilde{\rm X}) \ra \oh({\rm X}^*) + \h(^2{\rm S}_{1/2})
\hspace{0.5in} ({\rm Branch\; 2a, 66\%}), 
\ee
\be
\label{h2o[hdiss2b}
h\nu + \htwoo(\tilde{\rm X}) \ra \oh(\rm A) + \h(^2{\rm S}_{1/2})
\hspace{0.5in} ({\rm Branch\; 2b, 13\%}), 	
\ee
\be
\label{h2o[hdiss2c}
h\nu + \htwoo(\tilde{\rm X}) \ra {\rm O}(^3{\rm P}) + 2\h(^2{\rm S}_{1/2})
\hspace{0.5in} ({\rm Branch\; 2c, 21\%}),   
\ee
where $\oh(\rm X^{*})$ here indicates that OH is left in highly excited levels even for
$v''= 0$ and 1. We ignore any possible contribution of this energy to heating.

We first estimate the direct heating from product kinetic energy for the four ways that 
$\htwoo$ photodissociates. For the relatively soft photons of Band 1, the effective photons have a mean energy of 7.5\,eV, and with $D(\htwoo) = 5.10$\,eV, 2.42\,eV is available for heating. Using the results in  Andressan et al.~(1986), 88\% goes into  kinetic energy and the direct heating from Band 1 is, 
\be
\label{heat1direct}
Q_{\rm dir}({\rm Band\,1}) = 0.88 \, [ h\bar{\nu} - D(\htwoo) ] \simeq 2.14\, {\rm eV}.
\ee

For Branch 2a, the OH X ground level is left in highly excited rotational and vibrational states.
We assume conservatively that this involves the energy equivalent of the excitation of the OH\,A 
level, 4.0\,eV. Because we also assume that the critical density of the OH X transitions is very high, 
little of the high OH\,X excitation energy is available for heating. The mean photon energy of 
the broad Band 2 cross section from 1100 to 1450 \AA \,is 9.67\,eV, and a rough estimate of the available
energy is, 
\be
\label{heat2achem}
\Delta E_{\rm chem}({\rm Band\,2a}) \simeq  h\bar{\nu} - D(\htwoo) - E({\rm A-X})
\simeq 0.57\, {\rm eV}.
\ee
About one-half of this is available for direct heating.
The estimate for Branch 2b is basically the same because this branch leads to OH in its first 
electronic level\,A, so Eq.~\ref{heat2achem} is a good estimate for this case. The excitation 
in this case is lost by fluorescence near 3000\,\AA \, because the OH\,A level requires 
very high densities for de-excitation. Lastly, Branch 2c leads to the ground level of the O\,I 
atom and requires 9.51\,eV to completely dissociate $\htwoo$. Recalling that the mean 
photon energy is 9.67\,eV, only $\simeq 0.16$\,eV is available for heating and excitation. Averaging 
over the branching ratios in Eqs.~\ref{h2o[hdiss2a}, \ref{h2o[hdiss2b} and \ref{h2o[hdiss2c} and assuming
one-half of this is available for direct heating, the average direct heating from Band 2 is
{$\simeq 0.24$\,eV.

All of the $\htwoo$ photodissociation pathways can generate chemical heating initiated by the 
O and OH reactions~\ref{R:O+H2} and \ref{R:OH+H2} in warm regions and discussed in Sec.~3.3 
on CO. Reactions~\ref{R:O+H2} and \ref{R:OH+H2} involve chemical energy changes of 0.63\,eV and 0.56\,eV, 
respectively, and roughly one-half might go directly into heating in moderately dense regions, while the 
full amounts might be available in very dense regions. All of these considerations are summarized by 
the estimates in Table 4. The chemical heating in columns 3 and 4 is based on the assumption that 
one-half of the chemical energy associated with the formation of OH and O and their subsequent reactions
goes into heating. It is noteworthy that the photochemical heating from Bands 1 and 2 are rather different.
Inspection of Eq.~\ref{h2o[hdiss1} with Eqs.~\ref{h2o[hdiss2a}, \ref{h2o[hdiss2b} and \ref{h2o[hdiss2c} 
shows that the outcomes from Branch 2 involve higher energy products from which it is more difficult to extract 
heating as discussed above. The average value for Branch 2 is calculated by a simple average over 2a, 2b and 2c 
using the branching rations in Eqs.~\ref{h2o[hdiss2a}, \ref{h2o[hdiss2b} and \ref{h2o[hdiss2c}.}
} 
\begin{center}
Table 4. Photochemical Heating of $\htwoo^1$	\\	
\begin{tabular}{lccccc}    
\hline
Branch 		& Direct 	& Chemical  	& Chemical 			& Total	Heating	&  $N(\h)$   	\\
	 	&	  	& Moderately Dense	& Very Dense		& Very Dense	& 		 \\
\hline
1	  	& 2.14 		& 0.60			& 0.91	 		& 3.05		& 2		\\
\hline
2 (average)	& 0.24		& 0.55			& 0.86			& 1.10		& 2.2	\\
\hline
%Total		& 2.38		& 1.15			& 1.77			& 4.15		& 4.2	\\
%\hline
\multicolumn{6}{l}{$^1$Units are eV per photodissociation.}
\end{tabular}
\end{center}
To obtain these results, we assumed that the incident FUV field was constant over Bands 1 and 2. If that is not the case, 
a more accurate accounting can be obtained by carrying out a similar analysis on a smaller wavelength scale. 
Special attention may be required in treating the Ly$\alpha$ line, e.g., in the case of protoplanetary disks where it 
is very strong (e.g., Schindhelm et al.~2012). For densities intermediate between the limiting moderate and very dense 
cases used here for purposes of rough estimation, the level populations of the relevant species need to be calculated, 
including collisions as well as radiative decay. The last column of Table 4 indicates that a significant amount 
of the heating may arise from the conversion of atomic to molecular hydrogen that would be ordinarily included in $\hm$ 
formation heating.

\subsection{OH}

According to theoretical calculations by van Dishoeck \& Dalgarno (1984), there are four main channels for the photodissociation of OH  from 900 to 1900\,\AA, summarized in Table 5,
\begin{center}
Table 5. OH Photodissociation$^1$	\\
\begin{tabular}{ccccc}    
\hline
Branch & OH Level      & Product & Wavelength (\AA) & KE (eV) \\
\hline
1	& 1 $^2\Sigma$	& O($^3$P$_J$)	& 1300-2000	& 3.50		\\
2	& 1 $^2\Delta$	& O($^1$D$_2$)	& 1000-1500	& 3.45		\\
3	& 3 $^2 \Pi$	& O($^1$D$_2$)	& 1000-1300	& 5.28		\\
4	& B $^2\Sigma$	& O($^1$S$_0$)	& 900-1300	& 2.49		\\
%	& 	& 	& 	&		\\
\hline
\multicolumn{5}{l}{$^1$ van Dishoeck \& Dalgarno (1984)}
\end{tabular}
\end{center}
The second column lists the excited OH level and the third column the energy level of the product O atom. Shortward of 1400\,\AA, Channels 2-4 lead to excited levels of the oxygen atom, $^1$D$_2$ at 1.97\,eV and 
$^1$S$_0$ at 4.19\,eV (both measured from ground). The $^1$S$_0$ level quickly decays to the $^1$D$_2$ 
level which can be collisionally de-excited.
The fragment kinetic energy in the fifth column of the table is the average photon energy diminished by the sum of the OH dissociation energy 4.41\,eV and the O\,I excitation energies for branches 2-4. The calculations were done by dividing each branch into 100\,\AA \,intervals, assuming the FUV flux is constant for each branch. Because of shielding by $\hm$, C, CO, and $\ntwo$ from 900-1100\,\AA, the contributions from channels 3 and 4 may be significantly reduced. Of course most of the entire wavelength band from 900 to 1900\,\AA\,is shielded by dust and molecules such as water.

The O atoms may be collisionally de-excited and contribute to the heating at large densities. The $^1$D$_2$ level 
has relatively small A-values and is likely to be collisionally de-excited in dense regions with $\nh \gtrsim 10^{10} \pcc$, thus leading to $\sim 2.0$\,eV in heating. The $^1$S$_0$ level has a much higher critical density, $\nh \sim 10^{12} \pcc$, so emission of a 5577\,\AA\,photon is more likely than collisional de-excitation in many applications. In any case, channels 2-4 will all lead to heating $\sim 2.0$\,eV by collisional de-excitation of the $^1$D$_2$ level, ignoring the small heating from collisional de-excitation of the fine-structure levels of the ground level.

There will also be some chemical heating from the reactions of the product radicals O and H. The chemical 
heating for O has already been estimated in the discussion of the photochemical heating of CO following 
Eq.~\ref{Oheateq} and is 0.6\,eV for all branches. The chemical heating for Branches 2-4 will be further 
enhanced by 2.0\,eV by the collisional de-excitation of the $^1$D$_2$ level. These estimates are summarized 
in Table 6.
\begin{center}
Table 6. Photochemical Heating of OH$^1$		
\begin{tabular}{lccccc}    
\hline
Branch 	& Direct 	& Chemical 	& Chemical	&Total$^2$	&  $N(\h)$  	 \\
	&	& Moderately Dense	& Very Dense	&Very Dense	& 	  	 	\\
\hline
1	& 3.5     	& 0.6			& 0.6	 	&4.1		& 3			\\
2	& 3.5     	& 0.6			& 2.6	 	&6.1		& 3			\\
3	& 5.3     	& 0.6			& 2.6	 	&7.9		& 3			\\
4	& 2.5     	& 0.6			& 2.6	 	&5.1		& 3			\\
\hline
\multicolumn{6}{l}{$^1$Units are eV per photodissociation}	\\
\multicolumn{6}{l}{$^2$ Total is the sum of direct plus chemical for the very dense case.}
\end{tabular}
\end{center}
Again there will be additional $\hm$ formation heating from the creation of 3\,H atoms. According 
to Table 6, OH is the exceptional case where the chemical heating does not dominate direct heating.

\section{Discussion}

Table 7 gives an overview of the estimates of photochemical heating made in the previous 
section. They have uncertainties that are a good fraction of an eV per photodissociation. 
The potentially significant variation of the incident FUV flux with wavelength has been 
ignored, and the rare case of dense gas with $\nh > 10^{11} \pcc$ has been omitted. 
The two cases considered here are extreme in the sense that one ({\it moderately dense}) 
is too low for extracting heat by collisional de-excitation and the other ({\it very dense}) 
is high enough for collisional de-excitation to be effective. For intermediate densities, 
the level populations of the relevant species need to be calculated, including the effects of 
collisions as well as radiative decay. In the case of atomic C, two values are given 
corresponding to the destruction of $\cp$ by reaction with $\hm$ and $\htwoo$ discussed in 
Sec.~3.  For OH a cross section-weighted average over the four branches in Table 6 was made.  
In the case of $\htwoo$, a simple average was taken of the two bands in Table 4. The last 
column gives the average number of H atoms they generate. The associated photochemical heating 
has been omitted  because it is usually included in $\hm$ formation heating.

\begin{center}
Table 7. Summary of the Average Photochemical Heating$^1$	\\	
\begin{tabular}{cccccc} 
\hline
Absorber 	& Direct 	& Chemical 		&  Chemical	&  Total$^2$	& $N(\h)$   	\\
	 	&	  	& Moderately Dense	& Very Dense	& Very Dense	& 		\\
\hline
C\;$^3$		& 1.1		& 3.5, 4.7		& 6.9, 9.5	& 8.0, 10.6	& 5		\\
$\hm$		& 1.2		& 1.2	 		& 11.3		& 12.5		& 2		\\
CO		& 1.4		& 3.6	 		& 7.3		&  8.7		& 3		\\
$\htwoo$	& 1.2		& 0.6 	 		& 0.9		&  2.1		& 2.1		\\
OH		& 3.6		& 0.6			& 1.9		&  5.5		& 3		\\
\hline
\multicolumn{6}{l}{$^1$ Units are eV per photodissociation.}			\\
\multicolumn{6}{l}{$^2$  Total is the sum of direct plus chemical for the very dense case.}			\\
\multicolumn{6}{l}{$^3$ The two values correspond to $\cp$ reactions with $\hm$ and $\htwoo$,
respectively.}	\\
\end{tabular}
\end{center}

Some rough generalizations are apparent from Table 7. The three species whose photo-destruction 
comes from the 912-1100\,\AA\, band, i.e., $\hm$, CO and C,  provide little direct heating but a 
large amount of chemical heating. By contrast the two species with smaller dissociation energies 
absorb the FUV band over a wide range of wavelengths and generate roughly comparable amounts 
of direct and chemical heating. All of the species considered generate a significant amount of total  
heating for each photodissociation. Because several are abundant in PDRs, photochemical heating has  
the potential to affect the thermal balance of dense molecular regions.   

An important conclusion to be drawn from these estimates of photochemical heating is that 
the results depend on density (as indicated by columns 3 and 4 in Table 7) and in some cases 
on temperature (as in the case of atomic C). The present estimates deal only with two 
extreme density regimes.  Because of the dependence on density and temperature, there is 
no single value for the photochemical heating, even for a given species, except in the special 
case of a completely uniform model of a molecular gas cloud. More generally, the photochemical 
heating needs to be calculated according to the local physical properties, with special attention 
given to the collisional de-excitation of the products of the initial photodissociation 
and the subsequent chemical reactions.  

Despite these warnings on the limitations of the estimates in Table 7, one can still get a rough 
idea about magnitudes in these particular cases by comparing the direct heating in column 2 and the 
estimated maximum heating for the very dense case in column 5. The direct heating for the five species 
are all between 1 and 2\,eV, except for the rather large value of 3.6\,eV for OH, where it tends to 
dominate the total heating. Thus the total heating for OH in the two limits are 4.2\,eV and 5.5\,eV, so 
adopting the average value of 4.8\,eV for the photochemical heating of OH will lead to errors of only 20\%. 
The water molecule is another special case because the chemical heating is not very different in the 
two limits in Table 7. Thus a good approximation for the total heating of $\htwoo$ is the average of 
1.8\,eV and 2.1\,eV for the moderately dense and the very dense limits, or 2.0\,eV. The situation for 
the three species whose dissociation occurs in the 912-1100\,\AA\, band is quite different because 
the direct heating is small and also because the chemical heating in the moderately dense and the 
very dense limits are rather different. The exception is CO where the average value of the total heating, 
6.8\,eV, is a fair approximation in both limits. The important case of $\hm$ and also atomic C have to 
be treated in the detail suggested in the previous paragraph. For applications where other species are 
of interest, the treatment of the five species in Table 7 should serve as a guide for obtaining 
preliminary estimates.  

The entries in column 5 of Table 7 for total heating of very dense regions of order 
10\,eV also illustrate the possibility of the total heating exceeding the initial photon energy.
This becomes clear when the energy gain from $\hm$ formation is considered. The photodissociation 
of CO provides a good example, where the dissociation products, C and O, react to reform CO 
(C + OH $\ra$ CO + H; Eq.~\ref{R:C+OH}) and $\htwoo$ (O + 2$\hm \ra \htwoo$ + 2H; Eq.~\ref{Oheateq}) 
and generate 3\,H atoms that can form $\hm$ molecules on dust grain surfaces. As discussed in Sec.~3.1.1 
after Table 2, the magnitude of $\hm$ formation heating is quite uncertain, but if we use our estimate 
of 1.3\,eV as an example of what might be recovered at very high densities, then the total photochemical 
heating of CO is $\sim 12.6$\,eV, close to the mean energy of the dissociating photons. The 
example of CO also shows how, following photodissociation CO, it may be quickly replaced by chemical reaction
of the product atomic C. When the results of Table 7 are included in a full-scale thermal-chemical model 
for protoplanetary disks, a steady state in temperature and chemical abundances is reached in which the 
net production and destruction of every species are balanced and the photochemical heating (and every 
other heating mechanism) is balanced by gas cooling processes.

One can also see from Table 7 that, with several of the total heating energies of order 10\,eV per photodissociation in dense regions, FUV photochemical heating dominates cosmic-ray heating. This is true 
even in only moderately dense regions. Although cosmic rays generate $\sim 10$\,eV per ionization (GGP12), 
typically there are one million fewer cosmic ray ionizations than FUV dissociations. The situation changes 
for thick clouds where the photo-transition region becomes optically thick to FUV radiation through 
absorption by dust and various neutral species.

More relevant than cosmic rays is the comparison between photochemical heating and photoelectric 
heating by dust and other small particles (e.g. PAHs),  generally believed to be the most important 
heating source for the ISM (Tielens 2005, Sec.~3.3.3; Draine 2010, Sec.~30.2). Both of these 
heating mechanisms are initiated by the broadband FUV flux and yield significant levels of 
heating per event, $\sim 2$\,eV for the case of photochemical heating of 
$\htwoo$ from Table 7 and $\sim 5$\,eV for photoelectric heating (Tielens 2005). Other difference can 
arise from the cross sections for the two processes and abundance factors.

For the photoelectric effect, we use the standard ISM dust cross section per H nucleus, 
$\sim 10^{-21} \sqcm$ (Spitzer 1978, Eq.~7-23) and an efficiency of $\sim 0.1$ for charged 
grains (Draine 2011, Sec.~30.2) and obtain an effective photoelectric cross-section for heating 
of $10^{-22} \sqcm$. For $\htwoo$ the mean cross section for Bands 1 and 2 is 
$ \sim 2.0 \times 10^{-18} \sqcm$. With an $\htwoo$ abundance of $\sim 10^{-4}$ for very dense 
regions, the effective cross section per H nucleus is then $\sim 2\times 10^{-22}\sqcm$. 
Taking into account the smaller heating per photodissociation from Table 7, $\htwoo$ photochemical 
heating can be comparable to photoelectric heating in very dense regions. This result is 
of course sensitive to assumptions about grain properties and the water abundance, so the 
present estimate may not apply in all cases. For example, for those moderately dense regions 
of the interstellar medium where the $\htwoo$ abundance is small, the photochemical heating will 
be reduced and grain photoelectric heating will dominate.

According to the above estimate, photochemical heating may well dominate photoelectric heating 
in the atmospheres of the inner regions of protoplanetary disks, where both the grain surface 
area and the PAH abundance are likely to be reduced, How these atmospheres are heated is a long 
standing issue (e.g., Glassgold, Najita \& Igea 2004, henceforth GNI04). Most modeling papers on protoplanetary disks consider a variety of heating mechanisms, and photoelectric heating often 
plays an important role (e.g., Gorti \& Hollenbach 204, 2009, 2011; Nomura et al.~2007; 
Woitke et al.~2009; Woods \& Willacy 2009: Bruderer, 2013). An exception is GNI04 and models derived therefrom, most recently AGN14. In this work, standard thin-disk heating, 
\be
\label{accheat1}
\Gamma_{\rm acc} = \frac{9}{4} \alpha_h \rho c^2 \Omega,
\ee
is generalized to three dimensions; $\Omega$ is the angular rotation speed, 
$\rho \approx (1+4\xhe) \nh \mh$ is the local mass density of the gas, $c$ is the local sound speed for gas temperature $T$ ($c^2 = kT/m$), and the mean particle mass is,
\be
m  = \frac{1+ 4\xhe}{1-\xhm +\xhe + \xel}\mh. 
\ee
Eq.~\ref{accheat1} can be written in a numerically useful form (units, erg\,$\pcc$\,$\ps$), 
\be
\label{accheat2}
\Gamma_{\rm acc} =  6.18\times10^{-23}\, \alpha_h \, 
(1-\xhm +\xhe + \xel) \, \nh \,T 
\, (\frac{M}{\Msun})^{1/2} 
(\frac{\mathrm{AU}}{r})^{3/2}.
\ee
$\alpha_h$ is a 3-d phenomenological parameter  of order unity in the surface of the disk, in contrast to the usual average value of the standard $\alpha$-disk parameter $ \sim 0.01$. In addition to the arguments given in Sec.~3.2 of GNI04, this formulation of accretion heating is supported by recent simulations of the magneto-centrifugal instability  by Hirose \& Turner (2011) and Bai \& Stone (2013). 

For purposes of comparison with photochemical heating, we evaluate Eq.~\ref{accheat2} 
for a T Tauri star with mass $0.5 \Msun$ and inner disk-atmosphere physical conditions corresponding to the surface region just below the transition between the upper hot atomic 
and the warm molecular regions: $T = 800$\,K, $\nh = 10^{10} \pcc$ and $\alpha_h = 0.5$, 
following AGN14. The accretion heating rate per unit volume in this case is then, 
\be
\label{1accheat}
\Gamma_{\rm acc} =  1.08\times10^{-20}\, \ergps \,
(\frac{\alpha_h}{ 0.5}) \, 
\nh \, (\frac{T}{800\,{\rm K}}) \,
(\frac{\mathrm{AU}}{r})^{3/2}.
\ee
This may be compared with the photochemical heating rate for $\htwoo$ based on 
Eq.~\ref{heating_general} for the same conditions at 1\,AU and an $\htwoo$ abundance 
of $ 10^{-4}$, 
\be
\label{waterphotochemheat}
\Gamma_{\rm phchem}(\htwoo) = 1.5\times10^{-20}\, \ergps\, \nh.
\ee  
Thus at 1\,AU at the top of the molecular region, the photochemical heating rate from $\htwoo$ 
is estimated to be about the same as the accretion heating in Eq.~\ref{1accheat}. With increasing depth 
the dissociating radiation rapidly decreases and accretion heating will dominate. Nonetheless, 
before strong water self-shielding sets in, the molecular transition region may be dominated by 
FUV photochemical heating. Photochemical heating is sensitive to the FUV flux level as well as 
to absorption by dust and molecules. The above estimate was made with a T Tauri star FUV luminosity of 
$5 \times 10^{31} \ergps$. The atlas of {\it Hubble Space Telescope} FUV luminosities of T Tauri 
stars (Yang et al.~2012) shows luminosities that are much larger and mainly much smaller than our 
choice. Therefore, FUV heating may not be important for T Tauri stars with small FUV luminosities. 
Because FUV and accretion luminosities correlate, significant photochemical heating may only apply 
to high accretion sources.

\section{Conclusion}

The main conclusion of this paper is that heating from the absorption of FUV radiation by neutral 
species in dense molecular gas can be dominated by the energy yield from the chemical reactions induced 
by the products of photodissociation or photoionization. The total heating, referred to here as 
photochemical heating, consists of heating from the slowing down of the initial photo fragments 
(direct heating) and the energy released by the ensuing chemical reactions (chemical heating). 
The products from both processes are often excited and, if the density is high 
enough, some of the excitation energy can be converted into heating by collisional de-excitation 
and lead to a significant amount of heating. A good example is provided by $\hm$, where formation 
on grains produces rovibrationally excited molecules, and FUV absorption yields electronically 
excited molecules whose fluorescent decay leads to excitation of ro-vibrational levels.

When the reactions of the photo-fragments are considered, atomic H is a likely product, either 
because it is itself a fragment or another radical such as O or C generates it in reaction with $\hm$. 
We have {\it not} included the heating obtained when these atoms form $\hm$ on grains and then, 
either directly or after collisional de-excitation, contribute to the total photochemical heating 
in dense regions. $\hm$ formation heating is generally included in thermal-chemical modeling
of molecular gas. Instead we have listed the mean number of atoms in Tables 2, 4 and 7. For example, 
this number is listed in column 6 of summary Table 7, and can then be used to estimate the total heating if the 
formation heating per newly formed $\hm$ molecule is known.

The occurrence of collisional de-excitation places demands on modeling the excitation of the  
species involved in photochemical heating in applications where the gas density varies significantly.  
In principle, the excitation has to be explicitly calculated and radiative decay and collisional 
de-excitation treated on an equal footing. Unfortunately, the necessary rate coefficients  
are not always known. In this paper, directed mainly to demonstrating proof of principle, 
we made approximate estimates of the photochemical heating in two limits that are either 
dense enough or not dense enough for collisional de-excitation to play a significant role. Our 
estimates may also be uncertain because they are restricted to a limited set of chemical reactions.

Another important conclusion of this study is that the magnitude of the total photochemical 
heating can be significant in the sense that it is comparable to and even larger than other 
familiar heating processes. This result is well illustrated by the case of the dense upper 
atmospheres of the inner regions of protoplanetary disks. This example also shows how 
photochemical heating is affected by the attenuation of the FUV radiation. More detailed modeling 
of the role of photochemical heating in protoplanetary disk atmospheres is underway and will be 
reported in subsequent publications. \\

The authors are pleased to acknowledge support from NASA grant NNG06GF88G (Origins of Solar System). 
They are grateful for helpful discussions with Mat\'e \'Ad\'amkovics 
and for the helpful comments of the referee which improved the manuscript. 
  
\newpage
\section{Appendix}
\begin{center}
Table A1.{\bf Main Reactions} 	\\
\vspace{0.05in}
\begin{tabular}{ccccl}    
\hline
No.	& Reaction	& Rate Coefficient ($\ccps$)	&  $\Delta E_{\rm chem}$ (eV) & reference \\
\hline
%	& 									& 			&	\\
R1	& $\cp + \hm \ra \chp + \h$ 	& $7.5 \times 10^{-10}\,e^{-4620/T}$			& -0.35	& 1	\\
R2	& $\cp + \hm \ra \chtwop + h\nu$& $4.0 \times 10^{-16}$					& 4.47 	& 2	\\
R3	& $\cp + \htwoo \ra \hcop + \h$	& $2.1 \times 10^{-9}$					& 5.33	& 3	\\
R4	& $\chthreep + \el \ra {\rm products}$ &	$ 2.0 \times 10^{-5} T^{-0.61} $	& 3.24	& 4	\\
R5	& $\chthree + \hm \ra  \chfour +\h$ & $1.80 \times 10^{-21 } \, T^{2.88} \, e^{-4060/T}$ & 7.46	& 5	\\
R6a	& ${\rm O} + \chtwo \ra \co + 2\h$		& $2.0 \times 10^{-10}\, e^{-270/T} $	& 3.22	& 5	\\
R6b     & ${\rm O} + \chtwo \ra \co + \hm$		& $1.4 \times 10^{-10}\, e^{-270/T} $	& 7.73	& 5	\\
R7	& $\h + \h + {\rm Gr} \ra \hm + {\rm Gr}^{*}$	& $\frac{1}{2} n(\h)v(\h) \nd <\pi a^2> \epsilon 
	S(\Tg,\Td)$	& 4.48	& 6	\\
R8	& $\el + \hcop \ra \co + \h$	& $1.8 \times 10^{-5}\,T^{-0.79}$			& 7.48	& 7	\\		
R9	& ${\rm O} + \hm \ra \oh + \h$	& $6.34\times 10^{-12} \, e^{-4000/T} $			& 	& 	\\
	&				& $ + 1.46\times 10^{-9} \, e^{-9650/T} $		&-0.065	& 5	\\
R10	& $\oh + \hm \ra \htwoo + \h$	& $3.60\times 10^{-16} \, T^{1.52} \, e^{-1740/T} $	& 0.63	& 5	\\
R11	& ${\rm C} + \oh \ra \co + \h$	& $7.0 \times 10^{-11}$					& 6.48	& 8	\\
\hline
\multicolumn{5}{l}{References 1.  Herr\'aez-Aguilar, D. et al.~2014; 2. UMIST; 3. Anicich 1993; 
4. Thomas et al.~2012;} \\		
\multicolumn{5}{l}{5. Baulch et al.~2005; 6. Cazaux \& Tielens 2004, as discussed in detail by GMN09;} \\
\multicolumn{5}{l}{7. Herman et al.~2014; 8. Lin et al.~2008}
\end{tabular}
\end{center}

\newpage

Table A2 gives the chemical energies of the relevant species  in terms of the enthalpies $\Delta H$ at STP, 
i.e., 300 K. They can differ from ground state energies by as much as a few tenths of an eV. The standard 
unit, kcal/mole is used, with 1\,eV = 23.06 kcal/mole and 1 kcal/mole = 4.1854 kJ/mole. For the neutral species, a good reference is Table 3.1 of Baulch et al.~(2005, “Evaluated Kinetic Data for Combustion Modeling: Supplement II”).  Additional information is available at the NIST Chemical Kinetics website \footnote{http://kinetics.nist.gov/kinetics/welcome.jsp}.  For ions one needs ionization potentials and/or proton affinities $pa$. Much of this information is available in the NIST Chemistry WebBook\footnote{http://webbook.nist.gov/chemistry/}. A useful table for hydrocarbon ions is given by Anicich (1984). For example, the enthalpy of a molecular ion is, 
\be
\Delta H({\rm M}\hp) = \Delta H(\hp) + \Delta H({\rm M}) - pa({\rm M}).
\ee

\begin{center}
Table A2. Enthalpies (kcal/mole)	\\
\vspace{0.05in}
\begin{tabular}{lclc}    
\hline
Neutral	& Enthalpy	&Ion			&  Enthalpy \\
\hline
%		&		&			&		\\
H		& 52.1	& $\hp$		& 366		\\		
H$_2$		& 0		& H$_2^+$		& 356		\\
		& 		& H$_3^+$		& 265		\\
C		& 171.3		& C$^+$		& 431		\\
O		& 59.5	& O$^+$		& 374		\\
CO	      &-26.4	& CO$^+$		& 297		\\
HCO		& 10.4	& HCO$^+$		& 198		\\
H$_2$CO   	& -24.9	& H$_2$CO$^+$	& 225		\\  
OH		& 8.9	& OH$^+$       	& 360		\\
H$_2$O      & -57.8	& OH$_2^+$		& 233		\\
		&		& H$_3$O$^+$	& 145		\\
O$_2$		& 0		& O$_2^+$		& 278		\\
O$_2$H	& 0.5	& O$_2$H$^+$	& 262		\\
CO$_2$ 	& -94.0	& CO$_2^+$		& 231		\\
O$_3$		& 34.1	& O$_3^+$		& 323		\\
CH		& 142.0		& CH$^+$		& 387		\\
CH$_2$	& 92.4	& CH$_2^+$		& 328		\\
CH$_3$	& 34.8	& CH$_3^+$		& 262		\\
CH$_4$ 	& -17.9	& CH$_4^+$		& 273		\\
		&		& CH$_5^+$		& 218		\\
\hline
%\multicolumn{3}{l}{Herr\'aez-Aguilar, D. et al.~2014}; 2.
\end{tabular}
\end{center}


\newpage

\begin{references}
 
\reference{} \'Ad\'amkovics, M., Glassgold, A.E. \& Najita, J.R. 2014, ApJ, 786:135 (AGN14)

\reference{} Anicich, V.G. et al. 1984, J. Phys. Chem., 88, No. 20

\reference{} Bechellerie, D. et al. 2009, Phys. Chem. Chem. Phys., 11, 2715

\reference{} Bai, X.-N. \& Stone, J.M. 2013, ApJ, 767:30 (18pp)

\reference{} D.L. Baulch et al. 2005, J. Phys. Chem. Ref. Data, Vol. 34, No. 3 

\reference{} Bruderer, S. 2013, A\&A, 559, A46

\reference{} Canto, A.M., Mazzini, M., Pettini, M. \& Tozzi, G.P. 1981, Phys. Rev. A23, 1223

\reference{} Cazaux, S. \& Tielens 2004, A. G. G. M., ApJ, 604, 222

\reference{} Clavel, J., Viala, Y.P. \& Bel, N, 1978, A\&A, 65, 435   

\reference{} Crovisier, J. 1984, A\&A, 130, 361

\reference{} Dalgarno, A. \& Oppenheimer, M. 1974, ApJ, 192, 597

\reference{} Dalgarno, A., Yan, M. \& Liu, W.-H. 1999, ApJS, 125, 237 

\reference{} Draine, B. 2011, {\it Physics of the Interstellar Medium} (Princeton)

\reference{} Draine, B. \& Bertoldi, F. 1996, ApJ, 468, 269 

\reference{} Du, F. \& Bergin, E. 2014, arXiv 1408:2026. 

\reference{} Fillion, J.H. et al. 2001, JCP, A105, 11414

\reference{} France, K. Yang, Hao \& Linsky, J. 2011, ApJ, 729:7

\reference{} France, K. et al. 2012, ApJ, 744:22

\reference{} Glassgold, A.E. \& Langer, W.D. 1973, ApJ, 179, L147

\reference{} Glassgold, A.E., Meijerink, R. \& Najita, J.R. 2009, ApJ. 701:143 (GMN09)

\reference{} Glassgold, A.E., Najita, J.R. \& Igea, J. 2004, ApJ, 615, 972 (GNI04)

\reference{} Glassgold, A.E., Galli. D. \& Padovani 2012, ApJ, 756:157

\reference{} Gorti, U. \& Hollenbach, D. 2004, ApJ, 613, 424

\reference{} Gorti, U. \& Hollenbach, D. 2009, ApJ, 690, 1539

\reference{} Gorti, U. \& Hollenbach, D. 2011, ApJ, 735:90 (20pp)

\reference{} Harich, S.A. et al. 2000, JCP, 113, 10073 

\reference{} Herman, R. et al. 2014, J. Phys. Chem., 118, 6034

\reference{} Herr\'aez-Aguilar, D. et al.~2014, Phys. Chem. Chem, Phys. 20, 24800

\reference{} Henry, J.W.H. \& McElroy, M.B. 1969, J. Atmos. Sci. 26, 912

\reference{} Ingleby, L. et al. 2009, ApJ, 703, L37

\reference{} Ip, W.-H. 1983, ApJ, 264,726

\reference{} Kami\'nska, M, et al. 2012,Phys. Rev. A81, 062701 

\reference{} Le Bourlot, J, Pineau des For\^ets, G. \& Flower, D.R. 1999, MNRAS, 305, 802 

\reference{} Le Bourlot, J., Le Petit, F., Pinto. C, Roueff, E. \& Roy, F. 2012, A\&A, 541, A76

\reference{} Lee, L.C \& Suto, M. 1986, Chem Phys, 110, 161

\reference{} Lemaire, J.L. et al. 2010, ApJ, 725, L156

\reference{} Le Petit, F., Nemme, C., Le Bourlot, J. \& Roueff, E. 2006, ApJS, 164, 506 

\reference{} Lin, S.Y., Guo, H. \& Honvault, P. 2008, Chem. Phys. Letts. 450, 140

\reference{} Marconi, M.L. \& Mendis, D.A. 1982, ApJ, 260, 386

\reference{} Mota, R. 2005, Chemical Physics Letters 416, 152

\reference{} Nomura, H., Aikawa, Y., Tsujimoto, M. Nakagawa, Y. \& Millar, T.J. 2007, ApJ, 661, 344

\reference{} Parkinson, W.H \& Yoshino, K. 2003, Chem Phys, 294, 31

\reference{} Rodgers S.D. \& Charnley, S.B. 2002, MNRAS, 330, 660

\reference{} Rollig, M. et al. 2007, A\&A, 67, 187

\reference{} Schindhelm, E. et al. 2012, ApJ. 746:97

\reference{} Semaniak, J. et al. 2012, ApJ, 698, 886

\reference{} Spitzer, L. Jr. 1978 {\it Physical Processes in the Interstellar Medium}, Wiley, NY

\reference{} Tielens, A.G.G.M. 2005, {\it The Physics and Chemistry of the Interstellar Medium}, Cambridge 

\reference{} Thomas, D.R. et al. 2005, Phys Rev, A71, 032711

\reference{} Thomas, D.R. et al. 2012, ApJ, 758, 55

\reference{} UMIST, http://udfa.ajmarkwick.net

\reference{} van Dishoeck, E.F. \& Dalgarno, A. 1984, ApJ, 277, 576

\reference{} van Dishoeck, E.F. \& Black, J.H. 1988, ApJ, 334, 771

\reference{} Woods, P.M. \& Willacy, K. 2009, ApJ, 693, 1360

\reference{} Woitke, P., Kamp, I. \& Thi, W/-F. 2009, A\&A, 501, 383

\reference{} Yang, H, Herczeg, G.J., Linsky, J.L. et al. 2012, ApJ, 744:121(19pp)

\end{references}
\end{document}